\documentclass{llncs} 

\begin{document}
\title{A theory independent Curry-De Bruijn-Howard correspondence}
\author{Gilles Dowek}
\institute{INRIA, 23 avenue d'Italie, CS 81321, 75214 Paris Cedex 
13, France.
{\tt gilles.dowek@inria.fr}}

\date{}
\maketitle
\thispagestyle{empty}

Brouwer, Heyting, and Kolmogorov have proposed to define constructive
proofs as algorithms, for instance, a proof of $A \Rightarrow B$
as an algorithm taking proofs of $A$ as input and returning
proofs of $B$ as output. Curry, De Bruijn, and Howard have developed
this idea further. First, they have proposed to express these
algorithms in the lambda-calculus, writing for instance $\lambda f^{A
\Rightarrow A \Rightarrow B} \lambda x^A~(f~x~x)$ for the proof of
the proposition $(A \Rightarrow A \Rightarrow B) \Rightarrow A
\Rightarrow B$ taking a proof $f$ of $A \Rightarrow A \Rightarrow B$
and a proof $x$ of $A$ as input and returning the proof of $B$
obtained by applying $f$ to $x$ twice.  Then, they have remarked that,
as proofs of $A \Rightarrow B$ map proofs of $A$ to proofs of $B$,
their type $\mbox{\em proof}(A \Rightarrow B)$ is $\mbox{\em proof}(A)
\rightarrow \mbox{\em proof}(B)$. Thus the function {\em proof}
mapping propositions to the type of their proofs is a morphism
transforming the operation $\Rightarrow$ into the operation
$\rightarrow$. In the same way, this morphism transforms cut-reduction
in proofs into beta-reduction in lambda-terms.

This expression of proofs as lambda-terms has been extensively used in
proof processing systems: Automath, Nuprl, Coq, Elf, Agda, etc.
Lambda-calculus is a more compact representation of proofs, than
natural deduction or sequent calculus proof-trees. This representation
is convenient, for instance to store proofs on a disk and to
communicate them through a network.

This has lead to the development of several typed lambda-calculi:
Automath, the system F, the system F$\omega$, the lambda-Pi-calculus,
Martin-L\"of intuitionistic type theory, the Calculus of
Constructions, the Calculus of Inductive Constructions, etc. And we
may wonder why so many different calculi are needed.

In some cases, the differences in the lambda-calculi reflect
differences in the logic where proofs are expressed: some calculi, for
instance, express constructive proofs, others classical ones.  In
other cases, they reflect differences in the inductive rules used to
define proofs: some calculi are based on natural deduction, others on
sequent calculus. But most of the times, the differences reflect
differences in the theory where the proofs are expressed: arithmetic,
the theory of classes---a.k.a. second-order logic---, simple type
theory---a.k.a. higher-order logic---, predicative type theory, etc.

Instead of developing a customized typed lambda-calculus for each
specific theory, we may attempt to design a general parametric 
calculus that permits to express the proofs of any theory. This way,
the problem of expressing proofs in the lambda-calculus would be
completely separated from that of choosing a theory.

A way to do this is to start from the lambda-Pi-calculus, that is
designed to express proofs in minimal predicate logic and to define a
theory in an axiomatic way, declaring a variable, or a constant, for
each axiom. This is the approach of the {\em Logical framework}
\cite{LF}.  Yet, a limit of this approach is that the beta-reduction
is too weak in presence of axioms, and we need to add axiom-specific
proof-reduction rules, such as the rules of G\"odel system T for the
induction axiom, to emulate cut-reduction in specific theories.

We have proposed in \cite{CousineauDowek} a different approach, where
a theory is expressed, not with axioms, but with rewrite rules, as in
Deduction modulo \cite{DHK,DW}. This has lead to the {\em
lambda-Pi-calculus modulo}, and its implementation, the system {\em
Dedukti} \cite{Boespflug}.

Although it is just a proof-checker, Dedukti is a universal
proof-checker \cite{BCH}.  By choosing appropriate rewrite rules, the
lambda-Pi-calculus modulo can be parametrized to express proofs of any
theory that can be expressed in Deduction modulo, such as arithmetic,
the theory of classes, simple type theory, some versions of set
theory, etc. By choosing appropriate rewrite rules, the
lambda-Pi-calculus can also emulate the system F, the system
F$\omega$, the Calculus of Constructions \cite{CousineauDowek}, the
Calculus of Inductive Constructions \cite{CoqInE}, etc.  This has lead
to the development of systems to translate proofs from the system Coq
to Dedukti \cite{CoqInE} and from the system HOL to Dedukti
\cite{Assaf}.

This universal proof-checker opens new research directions that still
remain to be investigated. First, what happens if we prove the
proposition $A \Rightarrow B$ in a theory ${\cal T}_1$ and the
proposition $A$ in a theory ${\cal T}_2$? Is there a theory in which
we can deduce $B$?  Of course, if the theories ${\cal T}_1$ and ${\cal
  T}_2$ are incompatible---such as set theory with the axiom of choice
and set theory with the negation of the axiom of choice---, it makes
no sense to deduce $B$ anywhere. But, there are also cases where one
of the rewrite systems expressing ${\cal T}_1$ and ${\cal T}_2$ in the
lambda-Pi-calculus modulo is a subset of the other, or where the union
of these two systems defines a consistent theory, or where propositions
and proofs of one theory may be translated into the other, and in all
these cases, it makes sense to deduce $B$ from the proofs of $A
\Rightarrow B$ and $A$, even if these proofs have been developed in
different theories and different systems.

More generally, although most proof processing systems are based on
strong theories---simple type theory, the Calculus of Inductive
Constructions, etc.---we know that many proofs developed in these
systems use only a small part of this strength. Making explicit the
axioms or rewrite rules defining these theories permits to identify
which axiom, or which rule, is used in which proof, in a similar way
as we, more or less, know which part of informal mathematics depends
on the axiom of choice and which part does not. 

Such an analysis may be a first step towards the development of
libraries of proofs, where proofs would not be classified in function
of the system in which they have been developed, but in function of
the axioms and rules they use, i.e. to a true interoperability between
proof systems.


\begin{thebibliography}{99.}

\bibitem{Assaf} A. Assaf, {\em Translating HOL in the lambda-Pi-calculus 
modulo}, Master thesis, in preparation, 2012.

\bibitem{Boespflug} M. Boespflug, {\em Conception d'un noyau de
v\'erification de preuves pour le lambda-Pi-calcul modulo}, Doctoral
thesis, \'Ecole polytechnique, 2011.

\bibitem{BCH} M. Boespflug, Q. Carbonneaux, and O. Hermant, The
lambda-Pi calculus modulo as a universal proof language, {\em Second
International Workshop on Proof Exchange for Theorem Proving},
2012.

\bibitem{CoqInE} M. Boespflug and
G. Burel, CoqInE: Translating the Calculus of
inductive constructions into the lambda-Pi-calculus modulo,
{\em Second International Workshop on Proof Exchange for Theorem
Proving}, 2012.

\bibitem{CousineauDowek} D. Cousineau and G. Dowek, Embedding Pure
type systems in the lambda-Pi-calculus modulo, in S. Ronchi Della
Rocca, {\em Typed lambda calculi and applications}, Lecture Notes in
Computer Science 4583, Springer-Verlag, 2007, pp. 102-117.

\bibitem{DHK} G. Dowek, Th. Hardin, and C. Kirchner, Theorem proving
 modulo, {\em Journal of Automated Reasoning}, 31, 2003, pp. 33-72.

\bibitem{DW} G. Dowek and B. Werner, Proof normalization modulo, {\em
The Journal of Symbolic Logic}, 68, 4, 2003, pp. 1289-1316.

\bibitem{LF} R. Harper, F. Honsell, and G. Plotkin, A framework for
defining logics, {\em The Journal of the ACM}, 40, 1, 1993.
\end{thebibliography}
\end{document}